\documentclass[ reprint, superscriptaddress,
 amsmath,amssymb,
 aps,
 prl
]{revtex4-2}

\usepackage{graphicx}
\usepackage{overpic}
\usepackage{dcolumn}
\usepackage{bm}
\usepackage[hidelinks,colorlinks=true,linkcolor=blue,citecolor=blue]{hyperref}
\usepackage[mathlines]{lineno}

\usepackage{braket}
\usepackage{color}
\usepackage{float}

\begin{document}
\title{Solutions of the Schr\"{o}dinger equation for anisotropic dipole-dipole  interaction plus isotropic  van der Waals interaction}
\author{Ruijie Du}

\affiliation{Department of Physics, Renmin University of China, Beijing, 100872,
China}

\author{Ran Qi}
\email{qiran@ruc.edu.cn}
\affiliation{Department of Physics, Renmin University of China, Beijing, 100872,
China}

\author{Peng Zhang}
\email{pengzhang@ruc.edu.cn}


\affiliation{Department of Physics, Renmin University of China, Beijing, 100872,
China}

\affiliation{Beijing Computational Science Research Center, Beijing, 100193, China}

\date{\today}
\begin{abstract}
By generalizing Bo Gao's approach [Phys. Rev. A {\bf 58}, 1728 (1998)] for solving the Schr\"{o}dinger equation  for an isotropic van der Waals (vdW)  potential to the systems with a multi-scale anisotropic long-range interaction,
we derive the  solutions for the Schr\"{o}dinger equation for an anisotropic dipole-dipole interaction plus an isotropic attractive vdW potential, {\it i.e.}, ${C_d(1-3\cos^2\theta)}/{r^3}-{C_6}/{r^6}$, which is projected to the subspace with angular momentum $l\leq l_{\rm cut}$, with $l_{\rm cut}$ being an arbitrary angular-momentum cutoff. Here $\theta$ is the polar angle of the coordinate ${\bm r}$  and  $r=|{\bm r}|$.
The asymptotic behaviors of these solutions for $r\rightarrow 0$ and $r\rightarrow \infty$ are obtained. These results can be used in the research of collisions and chemical reactions between ultra-cold polar molecules in a static electric field.  Our approach to derive the solutions can be applied to the systems with a general long-range potential $\sum_{\lambda= 2}^{\lambda_{\rm max}} {V_\lambda(\theta,\varphi)}/{r^\lambda}$,  with $\varphi$ being the azimuthal angle of ${\bm r}$, and thus can be used in various problems on molecule-molecule interaction.

\end{abstract}
\maketitle

{\it 1. Introduction.} In recent years ultracold  gases
of various  polar molecules were  successfully realized by experimental groups  
\cite{K40Rb87, Rb87Cs133_1, Rb87Cs133_2, Na23K40, Na23Rb87, Na23Li6, Na23K39, Na23Cs133, JW2019_2, He2020, XY2021, JW2022, XY2022},
and ultracold molecule physics attracted more and more attention \cite{Jun2009, KK2010, NB2016, Jun2017_1, Jun2017_2}. 
For two ultracold molecules in the ground internal state with a weak static electric field along the $z$-axis,  the inter-molecule interaction can be described by an anisotropic model 
\begin{eqnarray}
V({\bm r})=\frac{C_d(1-3\cos^2\theta)}{r^3}-\frac{C_6}{r^6}\label{mod}
 \end{eqnarray}
 in the long-range region where the inter-molecule distance $r$ is larger than a ``short-range length scale" $r_\ast$, which is usually  dozens of Bohr radius \cite{Dulieu2013}.
Here ${\bm r}$ is the relative position of these two  molecules, $\theta$ is the angle between ${\bm r}$  and the $z$-axis, $r=|{\bm r}|$ and $C_{d,6}>0$. The first term of Eq.~(\ref{mod}) is the dipole-dipole interaction, with the intensity $C_d$ being proportional to the square of the electric field, and the second term is the isotropic attractive van der Waals (vdW) potential. 
In the short-range region (SRR) with $r\lesssim r_\ast$ the inter-molecular interaction becomes complicated and no longer has a simple analytical expression.
The model of Eq.~(\ref{mod}) is widely-used
in the studies of ultracold molecule physics   \cite{John2010_1, John2010_2, John2011, Robin2012, Dajun2018}.

In this work we derive the  solutions of the Schr\"odinger equation
 for the
 relative motion of two molecules
with  interaction  $V({\bm r})$ of Eq.~(\ref{mod}), {\it i.e.},
\begin{eqnarray}
\left[-\frac{\hbar^2\nabla_{\bm r}^2}{2\mu}+V({\bm r})\right]\Psi({\bm r})=\epsilon\Psi({\bm r}),\label{se1}
\end{eqnarray}
which is projected 
 to the subspace with $l=0,1,2,...,l_{\rm cut}$, 
 with respect to arbitrary non-zero energy $\epsilon$.
 Here $l$ is the quantum number of the angular momentum of the relative motion and the cutoff $l_{\rm cut}$ could be arbitrary non-negative integer, and $\mu$ is the reduced mass.
 Explicitly, 
we  analytically express a group of special
solutions of this  equation, as well as the asymptotic behaviors of these solutions in the limits $r\rightarrow 0$
and $r\rightarrow \infty$, as functions of a group of indexes. These indexes are solutions of an algebraic equation. Other solutions of Eq.~(\ref{se1})  can be expressed as  linear combinations of the special solutions we derived.

Our results are helpful for the studies  of collisions and chemical reactions between ultracold  polar molecules,
 especially the theoretical studies with quantum defect theory (QDT) approaches \cite{John1998, Paul2005, Bo2008, Bo2010, Paul2011, John2012, GR2012, Paul2013, John2020, Bo2020}, 
 where the physics in the SRR are described as  boundary conditions satisfied by the 
 solution of Eq.~(\ref{se1})
 at  $r=r_\ast$ or $r\rightarrow 0$. Using our results one can directly obtain the behaviors of the wave functions in the long-range limit $r\rightarrow \infty$ corresponding to these short-range boundary conditions, as well as the scattering amplitudes and reaction rates. 
 
 The solutions of the Schr\"{o}dinger equation for an isotropic vdW  potential ({\it i.e.}, $C_d=0$) were derived by Bo Gao \cite{Bo1998} in 1998. These solutions were used in many studies of ultracold atoms and 
ultracold molecules in the absence of electric field \cite{Bo1998_2, Paul2005, Bo2008, Bo2010, Bo2014, Bo2016_2, Li2017, Peng2017, JW2018, JW2019}, and were generalized to the systems with isotropic $1/r^3$ potential \cite{Bo1999}, isotropic $1/r^4$ potential \cite{Bo2013}, and isotropic multi-scale  potentials \cite{Bo2016}. Here we generalize  Bo Gao's approach to the systems with anisotropic interaction for the first time and obtain our above results.

 Moreover,  the method we developed can be applied to derive the solution of other Schr\"odinger equation with a general long-range anisotropic potential  $\sum_{\lambda= 2}^{\lambda_{\rm max}} {V_\lambda(\theta,\varphi)}/{r^\lambda}$, with $\varphi$ being the azimuthal angle of ${\bm r}$. Since this type of potential is used to describe the  interaction between various kinds of molecules or atoms, our method can be applied to many problems of atom and molecule physics.
 
 In the following, we show the solutions we derived as well as the asymptotic behaviors. The derivation of these results and the application of our method to more general cases are illustrated in the supplementary material (SM) \cite{SM}.

{\it 2. The radial Schr\"odinger equaiton.} 
For our system, the angular momentum of the inter-molecular relative motion along the $z$-axis ($L_z$)
 and the spatial parity are conserved.  
For the cases with $L_z=m\hbar$ ($m=0,\pm 1, \pm 2,...$) and the spatial parity ${\rm P}$ (${\rm P}=\pm 1$ for even and odd parity, respectively), the solution $\Psi({\bm r})$ of 
 the projection of Eq.~(\ref{se1})
in the subspace with $0\leq l\leq l_{\rm cut}$, can be expanded as 
\begin{eqnarray}
\Psi({\bm r})=\sum_{l=l_a,l_a+2,l_a+4,...,l_b}\frac{u_l(r)}{r}{Y}_l^m(\theta,\varphi),\label{pi}
\end{eqnarray}
where ${Y}_l^m(\theta,\varphi)$ is the spherical harmonics and $l_{a,b}$ are defined as
\begin{eqnarray}
l_a&=&|m|\delta_{(-1)^{|m|},{\rm P}}+\big[|m|+1\big]\delta_{(-1)^{|m|},-{\rm P}}\ \ \ ;\\
l_b&=&l_{\rm cut}\delta_{(-1)^{l_{\rm cut}},{\rm P}}+\big[l_{\rm cut}-1\big]\delta_{(-1)^{l_{\rm cut}},-{\rm P}}\ \ ,
\end{eqnarray}
with $\delta_{j,j^\prime}$ being the  Kronecker delta. 
Thus, the amount of the terms in the summation of Eq.~(\ref{pi}) is 
\begin{eqnarray}
N\equiv \left(l_b-l_a\right)/2+1.
\end{eqnarray}
Therefore, the wave function $\Psi({\bm r})$ of Eq.~(\ref{pi}) is determined by
the $N$-component radial wave function ${\bm u}(r)$ which is defined as
\begin{eqnarray}
{\bm u}(r)
\equiv
\begin{pmatrix}
u_{l_a}(r)\\
u_{l_a+2}(r)\\
u_{l_a+4}(r)\\
...\\
u_{l_b}(r)
\end{pmatrix}.\label{uc}
\end{eqnarray}
\begin{widetext}
Accordingly, for the cases with $L_z=m\hbar$ and spatial parity P,  the projection of Eq.~(\ref{se1})
in the subspace with $0\leq l\leq l_{\rm cut}$ 
 can be re-expressed as the radial Schr\"odinger equation for ${\bm u}(r)$:
\begin{eqnarray}
\left[\frac{\textrm{d}^2}{\textrm{d}r^2}-\frac{{\mathbb C}}{r^2}-\frac{\mathbb D}{r^3}+\frac{\beta_6^4}{r^6}{\mathbb I}+\bar{\epsilon}\hspace{0.05cm} {\mathbb I}\right]{\bm u}(r)=0,\label{se}
\end{eqnarray}
where  $\beta_6=({2\mu C_6}/{\hbar^2})^{1/4}$, $\bar{\epsilon}={2\mu\epsilon}/{\hbar^2}$, ${\mathbb I}$ is the $N\times N$ identical  matrix, and ${\mathbb C}$ and ${\mathbb D}$ are $N\times N$ matrixes with elements
\begin{eqnarray}
{\mathbb C}_{l,l^\prime}&=&l(l+1)\delta_{l,l^\prime};\\
{\mathbb D}_{l,l^\prime}&=&\frac{4\mu C_d}{\hbar^2}(-1)^{m+1}\sqrt{(2l+1)(2l^\prime+1)}
\begin{pmatrix}
l^\prime&2&l\\
m&0&-m
\end{pmatrix}
\begin{pmatrix}
l^\prime&2&l\\
0&0&0
\end{pmatrix},
\hspace{0.5cm}(l,l^\prime=l_a, l_a+2, l_a+4,...,l_b). \label{dllp}
\end{eqnarray}
Here we use $l$ ($l=l_a, l_a+2, l_a+4,...,l_b$) as the labels of the components of ${\bm u}(r)$ and the rows and columns of the $N\times N$ matrixes. In Eq.~(\ref{dllp})
 ${\mathbb D}_{l,l^\prime}$ is just the matrix element of the dipole-dipole interaction $C_d(1-3\cos^2\theta)/r^3$
in the basis of spherical harmonics.

{\it 3. Solutions of Eq.~(\ref{se}).} 
For arbitrary non-zero energy $\epsilon$, we find $2N$ special solutions of Eq.~(\ref{se}), which can be expressed as
\begin{eqnarray}
{\bm f}_{\epsilon}^{(j)}(r)&\equiv&\sum_{n=-\infty}^{\infty}{\bm b}_n(\nu_j)\sqrt{r}J_{\nu_j+n}(\sqrt{\bar{\epsilon}}r);\label{f}\\
{\bm g}_{\epsilon}^{(j)}(r)&\equiv&\sum_{n=-\infty}^{\infty}(-1)^n{\bm b}_n(\nu_j)\sqrt{r}J_{-\nu_j-n}(\sqrt{\bar{\epsilon}}r),\ \ (j=1,2,...,N).\label{g}
\end{eqnarray}
Here $J_\eta(z)$ is Bessel functions of the first kind. In this work the power function of a complex number is defined as $z^\eta=|z|e^{i\eta{\rm arg}[z]}$, with  ${\rm arg}[z]\in(-\pi,\pi]$ being the argument of $z$, and thus $\sqrt{\bar \epsilon}=i\sqrt{|\bar \epsilon|}$ for $\bar \epsilon<0$.

In the following we show the definition of the index $\nu_j$ ($j=1,..,N$) and the expression of the $N$-component vector ${\bm b}_n(\nu_j)$.

{\it 3.1. The indexes $\nu_{1,...,N}$.}
The index $\nu_j$ ($j=1,..,N$) is the root of the equation
\begin{eqnarray}
\det\left[{\mathcal M}(\nu)\right]=0,\label{dm}
\end{eqnarray}
where ${\mathcal M}(\nu)$ is a $4N\times 4N$ matrix defined by:
\begin{eqnarray}
{\mathcal M}(\nu)
\equiv\Delta_6^2 {\cal A}^{(+)}(\nu) {\mathcal Q}^{(+)}(\nu) {\cal F}(\nu+4){\cal A}^{(-)}(\nu+4)-{\cal F}^{-1}(\nu)-{\mathcal G}(\nu)+{\cal U}+\Delta_6^2 {\cal A}^{(-)}(\nu) {\mathcal Q}^{(-)}(\nu)  {\cal F}(\nu-4){\cal A}^{(+)}(\nu-4),\nonumber\\
\label{mn}
\end{eqnarray}
with 
\begin{eqnarray}
\Delta_6=\bar{\epsilon}^2 \beta_6^4/16.
\end{eqnarray}
Here  ${\cal A}^{(\pm)}(\nu)$, ${\cal F}(\nu)$, ${\mathcal G}(\nu)$, and ${\cal U}$ are $4N\times 4N$ matrixes that can be partitioned into sixteen $N\times N$ blocks:
\begin{eqnarray}
{\cal A}^{(+)}(\nu)=
\begin{pmatrix}
g_{-4}(\nu+7){\mathbb I}&0&g_{-2}(\nu+5){\mathbb I}&\hspace{0.1cm}-\frac{\sqrt{\bar{\epsilon}}}{2(\nu+4)\Delta_6}{\mathbb D}\vspace{0.2cm}\\
 0&g_{-4}(\nu+6){\mathbb I}&0&\hspace{0.1cm}g_{-2}(\nu+4){\mathbb I} \\
 0& 0& g_{-4}(\nu+5){\mathbb I}&\hspace{0.1cm}0\\
 0&0 &0 &\hspace{0.1cm}g_{-4}(\nu+4){\mathbb I}
\end{pmatrix};&\hspace{6.5cm}&\label{an}
\end{eqnarray}
\begin{eqnarray}
&&{\cal A}^{(-)}(\nu)=\begin{pmatrix}
g_4(\nu-1){\mathbb I}&\hspace{0.1cm}0 &0 &0 \\
0&\hspace{0.1cm}g_4(\nu-2){\mathbb I}& 0&0 \\
g_2(\nu-1){\mathbb I}&\hspace{0.1cm}0&g_4(\nu-3){\mathbb I}& 0\vspace{0.2cm}\\
-\frac{\sqrt{\bar{\epsilon}}}{2(\nu-1)\Delta_6}{\mathbb D}&\hspace{0.1cm}g_2(\nu-2){\mathbb I}&0&g_4(\nu-4){\mathbb I}
\end{pmatrix};
\quad
{\cal F}(\nu)=\begin{pmatrix}
\frac{1}{(\nu+3)^2}{\mathbb I}&0 & 0&0 \\
 0&\frac{1}{(\nu+2)^2}{\mathbb I}&0 &0 \\
0 & 0&\frac{1}{(\nu+1)^2}{\mathbb I}& 0\\
 0&0 & 0&\frac{1}{\nu^2}{\mathbb I}
\end{pmatrix};\quad\label{bn}\\
\nonumber
\end{eqnarray}
\begin{eqnarray}
&&{\mathcal G}(\nu)=\Delta_6
\begin{pmatrix}
g_0(\nu+3){\mathbb I}&
-\frac{\sqrt{\bar{\epsilon}}}{2(\nu+2)\Delta_6}{\mathbb D}
&g_2(\nu+1){\mathbb I}&0\vspace{0.2cm}\\
-\frac{\sqrt{\bar{\epsilon}}}{2(\nu+3)\Delta_6}{\mathbb D}&g_0(\nu+2){\mathbb I}&
-\frac{\sqrt{\bar{\epsilon}}}{2(\nu+1)\Delta_6}{\mathbb D}
&g_2(\nu){\mathbb I} \vspace{0.2cm}\\
g_{-2}(\nu+3){\mathbb I}&
-\frac{\sqrt{\bar{\epsilon}}}{2(\nu+2)\Delta_6}{\mathbb D}
&g_0(\nu+1){\mathbb I}&-\frac{\sqrt{\bar{\epsilon}}}{2\nu\Delta_6}{\mathbb D}\vspace{0.2cm}\\
0&g_{-2}(\nu+2){\mathbb I}&
-\frac{\sqrt{\bar{\epsilon}}}{2(\nu+1)\Delta_6}{\mathbb D}
&g_0(\nu){\mathbb I}
\end{pmatrix};
\quad
{\cal U}=\begin{pmatrix}
{\mathbb C}+{\mathbb I}/4&0&0&0\\
0&{\mathbb C}+{\mathbb I}/4&0&0 \\
0&0&{\mathbb C}+{\mathbb I}/4&0\\
0&0&0&{\mathbb C}+{\mathbb I}/4
\end{pmatrix};\nonumber\\
\label{gn}
\end{eqnarray}
with the functions $g_{0,\pm 2,\pm 4}(x)$ being defined as
$g_4(z)={1}/[{z(z+1)(z+2)(z+3)}]$; $g_{-4}(z)=g_4(-z)$;  
$g_2(z)={4}/[{(z-1)z(z+1)(z+3)}]$;   $g_{-2}(z)=g_2(-z)$, and
$g_0(z)={6}/[{(z-2)(z-1)(z+1)(z+2)}]$.
Moreover, ${\mathcal Q}^{(\pm)}(\nu)$ in Eq.~(\ref{mn}) are $4N\times 4N$ matrixes and are given by the continued-fraction-like  recursion equation
 \begin{eqnarray}
{\mathcal Q}^{(\pm)}(\nu)&=&\frac{1}{1+{\cal F}(\nu\pm4)\big[{\cal G}(\nu\pm 4)-{\cal U}\big]-\Delta_6^2 {\cal F}(\nu\pm4) {\cal A}^{(\pm)}(\nu\pm4){\mathcal Q}^{(\pm)}(\nu\pm4){\cal F}(\nu\pm8){\cal A}^{(\mp)}(\nu\pm8)},
\label{qp}
\end{eqnarray}
where $\frac{1}{\cal T}$ means the inverse of the $4N\times 4N$ matrix ${\cal T}$.
Eq.~(\ref{qp}) implies 
\begin{eqnarray}
\lim_{z\rightarrow +\infty}{\mathcal Q}^{(+)}(\nu+z)=\lim_{z\rightarrow -\infty}{\mathcal Q}^{(-)}(\nu+z)={\cal I},
\label{qpcm}
\end{eqnarray}
with ${\cal I}$ being the $4N\times 4N$ identity matrix. Eq.~(\ref{qpcm}) consists with the requirement of the convergence of the summation in Eqs.~(\ref{f}, \ref{g}).
For any fixed  $\nu$, one can derive ${\mathcal Q}^{(\pm)}(\nu)$ via a recursion calculation based on
Eqs.~(\ref{qp}, \ref{qpcm}). Moreover, the number of the  recursion steps, which are required to derive a converged result for ${\cal Q}^{(\pm)}(\nu)$, almost {\it does not increase} with the momentum cut off $l_{\rm cut}$ \cite{lcut}.

Here we also emphasis that, if $\nu$ is a solution of Eq.~(\ref{dm}), then $-\nu$, $\nu^*$ and $\nu+n$ ($n=0,\pm 1, \pm 2,...$) are also solutions of this equation \cite{SM}. However, if $\nu$ is already chosen as one of the indexes $\nu_{1,2,...,N}$, then $-\nu$, $\nu^*$ and $\nu+n$ ($n=0,\pm 1, \pm 2,...$) cannot also be chosen as the indexes. Namely, if $\nu\in\{\nu_1,\nu_2,...,\nu_N\}$, then we must have $-\nu\notin\{\nu_1,\nu_2,...,\nu_N\}$, $\nu^\ast\notin\{\nu_1,\nu_2,...,\nu_N\}$ and $(\nu+n)\notin\{\nu_1,\nu_2,...,\nu_N\}$ ($n=0,\pm 1, \pm 2,...$).

In the calculations we can choose the indexes $\nu_j$ ($j=1,2,...,N$) from the roots of Eq.~(\ref{dm})
which satisfy ${\rm Re}[\nu_j]\geq 0$ and ${\rm Im}[\nu_j]\geq 0$, and satisfy $\lim_{{\bar\epsilon}\rightarrow 0^+}\nu_j=\lim_{{\bar\epsilon}\rightarrow 0^-}\nu_j=l_a+2j-3/2$ \cite{detm}. As in the cases with only an isotropic van der Waals potential \cite{Bo1998}, the indexes $\nu_j$ ($j=1,2,...,N$) are real only when $|{\bar \epsilon}|$ is small. In Fig.~\ref{fig1} (a) we show the  real and imaginary parts of the indexes for some typical cases with $m=0$, ${\rm P}=1$ ({\it i.e.}, even spatial parity) and $l_{\rm cut}=4$.

{\it 3.2. The vector  ${\bm b}_n(\nu_j)$.} To show the expression of the $N$-component vector ${\bm b}_n(\nu_j)$ ($n=0,\pm 1, \pm 2,...$) in Eqs.~(\ref{f}, \ref{g}), 
we define the $4N\times4N$ matrixes ${\cal S}^{(\pm)}_\beta(\nu)$ and $4N$-component vectors ${\bm B}_\alpha(\nu_j)$ as 
\begin{eqnarray}
{\cal S}_\xi^{(\pm)}(\nu)&\equiv&-\Delta_6{\cal Q}^{(\pm)}(\nu\pm4\xi\mp4){\cal F}(\nu\pm 4\xi){\cal A}^{(\mp)}(\nu\pm4\xi);\ \ (\xi=1,2,...),\label{smm}
\end{eqnarray}
 and  
\begin{eqnarray}
{\bm B}_\alpha(\nu_j)&\equiv&{\cal S}^{(\sigma_\alpha)}_{|\alpha|}(\nu_j){\cal S}^{(\sigma_\alpha)}_{|\alpha|-1}(\nu_j)...{\cal S}^{(\sigma_\alpha)}_{1}(\nu_j){\bm B}_0(\nu_j),\hspace{0.4cm}(j=1,...,N;\ \alpha=0,\pm 1,\pm 2,...),\label{bq}
\end{eqnarray}
respectively, with $\sigma_\alpha=+(-)$ for $\alpha\ge 0$ ($\alpha<0$), and ${\bm B}_0(\nu_j)$ satisfying
\begin{eqnarray}
{\mathcal M}(\nu_j){\bm B}_0(\nu_j)=0.
\label{mb}
\end{eqnarray}
In specific calculations, one can normalize ${\bm B}_0(\nu_j)$ according to the convenience. 

The $N$-component vector ${\bm b}_n(\nu_j)$ ($n=0,\pm 1,\pm 2,...$) in Eqs.~(\ref{f}, \ref{g}) are determined by the above $4N$-component vectors ${\bm B}_\alpha(\nu_j)$    via the relation
\begin{eqnarray}
{\bm B}_\alpha(\nu_j)\equiv
\begin{bmatrix}
{\bm b}_{4\alpha+3}(\nu_j)\\
 {\bm b}_{4\alpha+2}(\nu_j) \\
{\bm b}_{4\alpha+1}(\nu_j)\\	
{\bm b}_{4\alpha}(\nu_j)
\end{bmatrix},
\ \ (j=1,...,N;\ \alpha=0,\pm 1,\pm 2,...). \label{bigb}
\end{eqnarray}

\begin{figure}[t]
\begin{centering}
\begin{overpic}[width=0.32\textwidth]{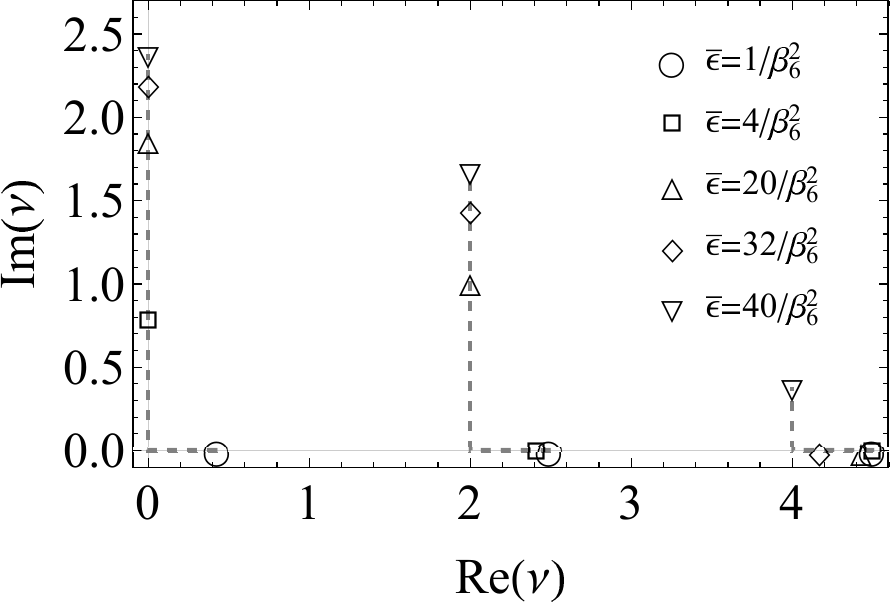}
\put(20,60){{\bf (a)}}
\end{overpic}
\begin{overpic}[width=0.3\textwidth]{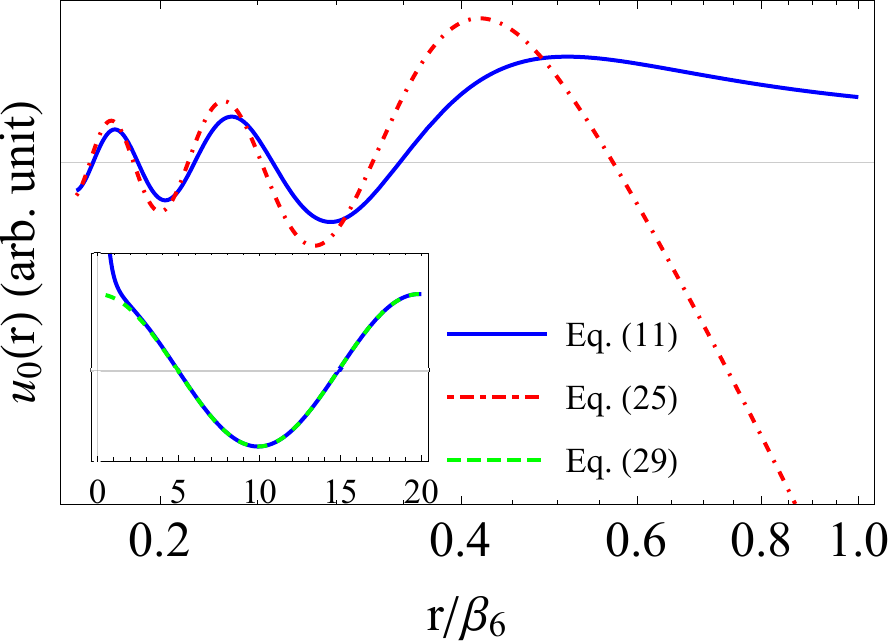}
\put(10,64){{\bf (b)}}
\end{overpic}\vspace{0.2cm}\\
\hspace{0.3cm}
\begin{overpic}[width=0.3\textwidth]{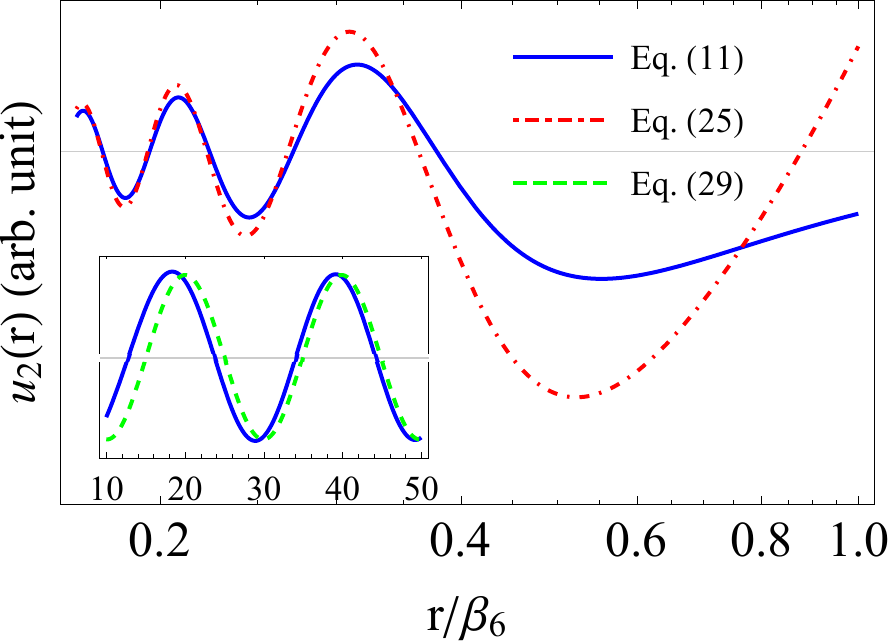}
\put(10,64){{\bf (c)}}
\end{overpic}
\begin{overpic}[width=0.3\textwidth]{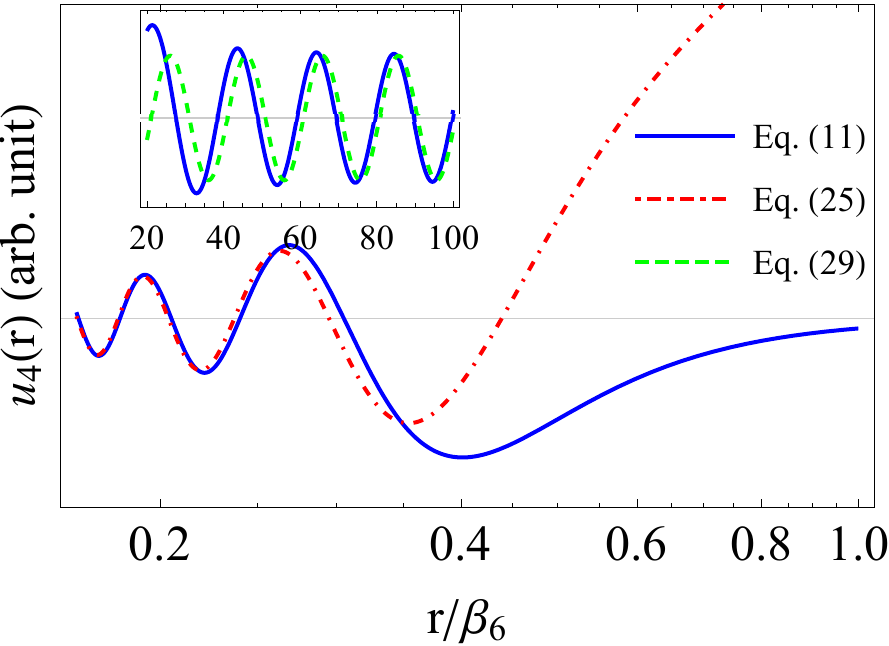}
\put(86,64){{\bf (d)}}
\end{overpic}
\par\end{centering}
\caption{ (color online) {\bf (a)}: The  indexes for the cases with $C_d=\hbar^2\beta_6/(2\mu)$, $m=0$, ${\rm P}=1$ ({\it i.e.}, even spatial parity), and $l_{\rm cut}=4$. In this case we have $l_a=0$, $l_b=4$, and $N=3$, and thus there are three indexes $\nu_{1,2,3}$. We show the position of these indexes in the complex plane.   {\bf (b-d)}: The radial wave functions $u_{l}(r)$ ($l=0,2,4$) of the solution ${\bm f}^{(1)}_{\epsilon}(r)$ given by Eq.~(\ref{f}) for the system of {\bf (a)} with $\epsilon=0.1\hbar^2/(2\mu\beta_6^2)$, and the corresponding asymptotic behaviors given by Eq.~(\ref{f0}) and Eq.~(\ref{fi1}) for $r\rightarrow 0$ and $r\rightarrow \infty$, respectively.}
\label{fig1}
\end{figure}

{\it 4. Asymptotic behaviors for $r\rightarrow 0$.} The behaviors of the solutions ${\bm f}_{\epsilon}^{(j)}(r)$ and ${\bm g}_{\epsilon}^{(j)}(r)$ ($j=1,...,N$) limit $r\rightarrow 0$ can be expressed as
\begin{eqnarray}
{\bm f}_{\epsilon}^{(j)}(r\rightarrow0)
&\rightarrow&\frac{2r^{3/2}}{\sqrt{\pi}\beta_6}
\left[{\bm p}_{{\rm c}-}(\nu_j)\cos\left(\frac{\beta_6^2}{2r^2}-\frac{\pi}{4}\right)+{\bm p}_{{\rm s}-}(\nu_j)\sin\left(\frac{\beta_6^2}{2r^2}-\frac{\pi}{4}\right)\right];\label{f0}\\
{\bm g}_{\epsilon}^{(j)}(r\rightarrow0)
&\rightarrow&\frac{2r^{3/2}}{\sqrt{\pi}\beta_6}
\left[{\bm p}_{{\rm c}+}(\nu_j)\cos\left(\frac{\beta_6^2}{2r^2}-\frac{\pi}{4}\right)+{\bm p}_{{\rm s}+}(\nu_j)\sin\left(\frac{\beta_6^2}{2r^2}-\frac{\pi}{4}\right)\right],\label{g0}
\end{eqnarray}
where the $N$-component vectors ${\bm p}_{{\rm c}\pm}(\nu_j)$  and ${\bm p}_{{\rm s}\pm}(\nu_j)$ are given by
\begin{eqnarray}
{\bm p}_{{\rm c}\pm}(\nu_j)\equiv\sum_{t=0,1,2,3}{\bm q}^{(\pm)}_t(\nu_j)\cos\left[\frac{\pi(\nu_j+t)}{4}\right];\hspace{0.6cm}
{\bm p}_{{\rm s}\pm}(\nu_j)\equiv\pm\sum_{t=0,1,2,3}{\bm q}^{(\pm)}_t(\nu_j)\sin\left[\frac{\pi(\nu_j+t)}{4}\right],\label{pcs}
\end{eqnarray}
with 
\begin{eqnarray}
{\bm q}^{(\pm)}_t(\nu_j)\equiv(\mp 1)^t\lim_{\alpha\rightarrow\pm\infty}\Bigg\{{\bm b}_{4\alpha+t}(\nu_j)\frac{\big({\rm-}16/\Delta_6\big)^{|\alpha|}\Gamma^2\big[|\alpha|+1\pm(\nu_j+t)/4\big]}{\big(\delta_\epsilon \Delta_6^{1/4}/2\big)^{\pm(\nu_j+t)}\Gamma(\mp\nu_j-4|\alpha|\mp t+1)}\Bigg\};\ \ (t=0,1,2,3).\label{qk}
\end{eqnarray}
Here $\delta_\epsilon=1(i)$ for $\epsilon>0$ ($\epsilon<0$) and $\Gamma(z)$ is the Euler's Gamma function.

{\it 5. Asymptotic behaviors for $r\rightarrow \infty$.}  
For $\epsilon>0$,
the behaviors of the solutions ${\bm f}_{\epsilon}^{(j)}(r)$ and ${\bm g}_{\epsilon}^{(j)}(r)$ ($j=1,...,N$) in the large-$r$ limit are given by
\begin{eqnarray}
&&{\bm f}_{\epsilon}^{(j)}(r\rightarrow \infty)\rightarrow\sqrt{\frac{2}{\pi\sqrt{\bar \epsilon}}}\bigg[{\bm d}_{{\rm c}-}(\nu_j)\cos\left(\sqrt{\bar \epsilon}r\right)-{\bm d}_{{\rm s}-}(\nu_j)\sin\left(\sqrt{\bar \epsilon}r\right)\bigg];\hspace{1cm} \text{(for $\epsilon>0$),}
\label{fi1}\\
&&{\bm g}_{\epsilon}^{(j)}(r\rightarrow \infty)
\rightarrow\sqrt{\frac{2}{\pi\sqrt{\bar \epsilon}}}\bigg[{\bm d}_{{\rm c}+}(\nu_j)\cos\left(\sqrt{\bar \epsilon}r\right)-{\bm d}_{{\rm s}+}(\nu_j)\sin\left(\sqrt{\bar \epsilon}r\right)\bigg];\hspace{1cm} \text{(for $\epsilon>0$).}
\label{gi1}
\end{eqnarray}
with 
\begin{eqnarray}
{\bm d}_{{\rm c}\pm}(\nu_j)&=&\sum_{n=-\infty}^{\infty}{\bm b}_n(\nu_j)
\cos\left(\pm\frac{\nu_j\pi}2-\frac{n\pi} 2-\frac{\pi} 4\right);\hspace{0.6cm}
{\bm d}_{{\rm s}\pm}(\nu_j)=\sum_{n=-\infty}^{\infty}{\bm b}_n(\nu_j)
\sin\left(\pm\frac{\nu_j\pi}2-\frac{n\pi} 2-\frac{\pi} 4\right).
\label{dcs}
\end{eqnarray}
For $\epsilon<0$, in the limit $r\rightarrow \infty$ the  solutions of Eq.~(\ref{f0})  exponentially increase or decay with $r$. A meaningful problem is that which solutions have the exponential-decay behavior.  We find that the $N$ solutions ${\bm h}^{(j)}_{\epsilon}(r)\equiv {\bm f}^{(j)}_{\epsilon}(r)-e^{i\nu_j\pi}{\bm g}^{(j)}_{\epsilon}(r)$ ($j=1,...,N$) satisfy
\begin{eqnarray}
{\bm h}^{(j)}_{\epsilon}(r\rightarrow\infty)\rightarrow-\sqrt{\frac{2}{\pi\sqrt{|\bar \epsilon|}}}e^{i\pi\nu_j/2}\sin(\pi\nu_j)\left[\sum_{n=-\infty}^{+\infty}(-i)^n{\bm b}_n(\nu_j)\right]e^{-\sqrt{|{\bar \epsilon}|}r}.\label{hrl}
\end{eqnarray}
Each solution of Eq.~(\ref{f0}), which exponentially decay to zero in the limit $r\rightarrow \infty$, is a linear combination of  ${\bm h}^{(j)}_{\epsilon}(r)$ ($j=1,...,N$).

As an example, in  Fig.~\ref{fig1} (b-d) 
 we show the exact radial wave functions $u_{l}(r)$ ($l=0,2,4$) of the solution ${\bm f}^{(1)}_{\epsilon}(r)$  obtained with our approach for the system of Fig.~\ref{fig1} (a) with $\epsilon=0.1\hbar^2/(2\mu\beta_6^2)$, as well as the corresponding asymptotic behaviors given by Eq.~(\ref{f0}) and Eq.~({\ref{fi1}}).

\end{widetext}

{\it 6. Other solutions.} 
It is difficult to analytically proof that the $2N$ solutions ${\bm f}_{\epsilon}^{(j)}(r)$ and ${\bm g}_{\epsilon}^{(j)}(r)$ ($j=1,...,N$) of  Eq.~(\ref{se}) are linearly independent. However, it is easy to numerically confirm this linear-independence property. Explicitly, these solutions are linearly independent when $\det[{\rm C}]\neq 0$, where 
\begin{eqnarray}
{\rm C}\equiv\left[
\begin{array}{l}
{\bm d}_{\rm c-}(\nu_1),...,{\bm d}_{\rm c-}(\nu_N),{\bm d}_{\rm c+}(\nu_1),...,{\bm d}_{\rm c+}(\nu_N)\\
{\bm d}_{\rm s-}(\nu_1),...,{\bm d}_{\rm s-}(\nu_N),{\bm d}_{\rm s+}(\nu_1),...,{\bm d}_{\rm s+}(\nu_N)
\end{array}
\right]\nonumber\\
\end{eqnarray}
is a $2N\times 2N$ matrix. So far we have not found any case where this condition is violated. Under the condition $\det[{\rm C}]\neq 0$, every solution of Eq.~(\ref{se})
can  be constructed as a linear combination of  ${\bm f}_{\epsilon}^{(j)}(r)$ and ${\bm g}_{\epsilon}^{(j)}(r)$ ($j=1,...,N$). As an example, in Sec.~C of our SM \cite{SM} we show the solutions with energy-independent asymptotic behaviors for $r\rightarrow 0$.


{\it 7. Application of our method for other problems.} 
As mentioned above, our method to derive the  solutions of Eq.~(\ref{se}) can be generalized to problems with  general potential $V({\bm r})=\sum_{\lambda=2}^{\lambda_{\rm max}} {V_\lambda(\theta,\varphi)}/{r^\lambda}$. 
Some details of this generalization are shown in
Sec.~D of our SM \cite{SM}.
Nevertheless, if the potential is still anisotropic in the limit $r\rightarrow 0$, 
the  asymptotic behavior of the solutions in this limit cannot be analytically obtained. One should derive these asymptotic behavior by 
 numerically fitting the special solutions of the Schr\"odinger equation, which are obtained via our approach,
 with the WKB solutions in the limit $r\rightarrow 0$.

This work is supported by National Key Research and Development Program of China Grant No. 2018YFA0306502 and NSAF Grant No. U1930201, No. 12022405 and No. 11734010, and the Beijing Natural Science Foundation (Grant No. Z180013). 
\bibliography{mqdt}

\begin{thebibliography}{48}
\bibitem [{lcu()}]{lcut}%
  \BibitemOpen
  \href@noop {} {}\bibinfo {note} {This can be explained with the following
  facts. In the recursion, ${\cal Q}^{(\pm)}(\nu)$ is expressed in terms of the
  matrixes ${\cal A}^{(\pm)}(\nu\pm 4n)$, ${\cal F}(\nu\pm 4n)$, ${\cal
  G}(\nu\pm 4n)$ ($n=0, 1, 2,...$), as well as the constant matrix ${\cal U}$.
  According to these definitions, the elements of matrixes ${\cal
  A}^{(\pm)}(\nu\pm 4n)$, ${\cal F}(\nu\pm 4n)$, ${\cal G}(\nu\pm 4n)$ all
  decay to zero in the limit $n\rightarrow+\infty$, and the decay speeds do not
  increase with $l_{\rm cut}$ or the matrix dimension.}\BibitemShut {Stop}%
\bibitem [{det()}]{detm}%
  \BibitemOpen
  \href@noop {} {}\bibinfo {note} {In the limit ${\bar\epsilon}\rightarrow 0$
  we have ${\cal M}(\nu)=-{\cal F}^{-1}(\nu)+{\cal U}$. Thus, in this limit we
  have ${\rm det}[{\cal M}(l_a+2i-3/2)]=0\ (i=1,...,N)$.}\BibitemShut {Stop}%
\end{thebibliography}
\global\long\def\id{\mathbbm{1}}
\global\long\def\ui{\mathbbm{i}}
\global\long\def\ud{\mathrm{d}}

\setcounter{equation}{0} \setcounter{figure}{0}
\setcounter{table}{0} 
\renewcommand{\theparagraph}{\bf}
\renewcommand{\thefigure}{S\arabic{figure}}
\renewcommand{\theequation}{S\arabic{equation}}

\onecolumngrid
\flushbottom

\section*{Supplementary Material}

\subsection*{A. Derivation of the Solutions ${\bm f}_{\epsilon}^{(j)}(r)$ and ${\bm g}_{\epsilon}^{(j)}(r)$}

In this section we derive the special solutions ${\bm f}_{\epsilon}^{(j)}(r)$
 and 
${\bm g}_{\epsilon}^{(j)}(r)$ ($j=1,...,N$) of Eq.~(\ref{se}). To this end,
we expand the solution ${\bm u}(r)$ of Eq.~(\ref{se})  as a
 Neumann series:
\begin{eqnarray}
{\bm u}(r)=\sqrt{r}\sum_{n=-\infty}^{\infty}{\bm b}_n(\nu)J_{\nu+n}(x),\label{uex}
\end{eqnarray}
where 
\begin{eqnarray}
x=\sqrt{\bar{\epsilon}}r,
\end{eqnarray}
and $\{J_{\nu+n}(x)|n=0,\pm 1, \pm 2,...\}$  is the first-kind Bessel function.
We substitute
 the form (\ref{uex}) into Eq.~(\ref{se}) and then obtain 
\begin{eqnarray}
\left(x^2\frac{\textrm{d}^2}{\textrm{d}x^2}{\mathbb I}+x\frac{\textrm{d}}{\textrm{d}x}{\mathbb I}+x^2{\mathbb I}-{\mathbb C}-\frac{1}{4}{\mathbb I}-\frac{\sqrt{\bar{\epsilon}}}{x}{\mathbb D}+\frac{16\Delta_6}{x^4}{\mathbb I}\right)\Bigg\{\sum_{n=-\infty}^{\infty}{\bm b}_n(\nu)J_{\nu+n}(x)\Bigg\}=0,\label{3e}
\end{eqnarray}
where $\Delta_6$ and the $N\times N$ matrixes ${\mathbb I}$, ${\mathbb D}$, and ${\mathbb C}$ are defined in our main text. Substituting the properties of Bessel function
\begin{eqnarray}
\left(x^2\frac{\textrm{d}^2}{\textrm{d}x^2}+x\frac{\textrm{d}}{\textrm{d}x}+x^2\right)J_{\eta}(x)&=&
\eta^2 J_{\eta}(x);\label{jp0}\\
\frac{2}{x}J_{\eta}(x)&=&\frac{1}{\eta}[J_{\eta+1}(x)+J_{\eta-1}(x)]\label{jp}
\end{eqnarray}
into Eq.~(\ref{3e}), we further derive the 
the recursion equation of ${\bm b}_n(\nu)$:
\begin{eqnarray}
&&\left[(\nu+n)^2{\mathbb I}-{\mathbb C}-\frac{1}{4}{\mathbb I}\right]{\bm b}_n(\nu)-\frac{\sqrt{\bar{\epsilon}}}{2(\nu+n+1)}{\mathbb D}{\bm b}_{n+1}(\nu)-\frac{\sqrt{\bar{\epsilon}}}{2(\nu+n-1)}{\mathbb D}{\bm b}_{n-1}(\nu)
+\Delta_6g_4(\nu+n-4){\bm b}_{n-4}(\nu)
\nonumber\\
&&+\Delta_6\Big[g_2(\nu+n-2){\bm b}_{n-2}(\nu)+g_0(\nu+n){\bm b}_n(\nu)+g_{-2}(\nu+n+2){\bm b}_{n+2}(\nu)+g_{-4}(\nu+n+4){\bm b}_{n+4}(\nu)\Big]\nonumber\\
&=&0,\label{rr}
\end{eqnarray}
with the functions $g_{0,\pm 2,\pm 4}(z)$ being defined in our main text.

In the recursion equation (\ref{rr}), ${\bm b}_{n}$ is related to not only ${\bm b}_{n\pm 1}$ but also ${\bm b}_{n\pm 2}$ and ${\bm b}_{n\pm 4}$. This fact inspires us to combine ${\bm b}_{n}$ and  ${\bm b}_{n+1}$, ${\bm b}_{n+2}$, ${\bm b}_{n+3}$ together, {\it i.e.}, introduce the $4N$-component vector  ${\bm B}_\alpha(\nu)$ which is defined as
\begin{eqnarray}
{\bm B}_\alpha(\nu)\equiv
\begin{bmatrix}
{\bm b}_{4\alpha+3}(\nu)\\
 {\bm b}_{4\alpha+2}(\nu) \\
{\bm b}_{4\alpha+1}(\nu)\\	
{\bm b}_{4\alpha}(\nu)
\end{bmatrix},
\ \ (\alpha=0,\pm 1,\pm 2,...).\label{bb}
\end{eqnarray}
 Accordingly, the recursion equation (\ref{rr}) of ${\bm b}_{n}(\nu)$ can be re-written as the one of ${\bm B}_\alpha(\nu)$:
\begin{eqnarray}
&&\Delta_6{\mathcal A}^{(+)}(\nu+4\alpha){\bm B}_{\alpha+1}(\nu)+\bigg[{\mathcal F}^{-1}(\nu+4\alpha)+{\mathcal G}(\nu+4\alpha)-{\cal U}\bigg]{\bm B}_\alpha(\nu)+\Delta_6 {\mathcal A}^{(-)}(\nu+4\alpha){\bm B}_{\alpha-1}(\nu)=0;\nonumber\\
&&\hspace{12cm}(\alpha=0,\pm 1, \pm 2,...),\label{rbb2}
\end{eqnarray}
where $4N\times 4N$ matrixes ${\cal F}$, ${\cal G}$ and ${\cal A}^{(\pm)}$ are defined in our main text.
We further formally express ${\bm B}_{\alpha}(\nu)$ ($\alpha=\pm 1, \pm 2,...$) as
\begin{eqnarray}
{\bm B}_\alpha(\nu)&=&{\cal S}^{(\sigma_\alpha)}_{|\alpha|}(\nu){\cal S}^{(\sigma_\alpha)}_{|\alpha|-1}(\nu)...{\cal S}^{(\sigma_\alpha)}_{1}(\nu){\bm B}_0(\nu);\hspace{0.4cm}(\alpha=\pm 1,\pm 2,...),\label{bqa}
\end{eqnarray}
with $\sigma_\alpha=+(-)$ for $\alpha> 0$ ($\alpha<0$). Substituting Eq.~(\ref{bqa}) into Eq.~(\ref{rbb2}), we  obtain the equations
\begin{eqnarray}
\left[\Delta_6{\mathcal A}^{(+)}(\nu+4\xi){\mathcal S}^{(+)}_{\xi+1}(\nu)+{\mathcal F}^{-1}(\nu+4\xi)+{\mathcal G}(\nu+4\xi)-{\cal U}+\Delta_6{\mathcal A}^{(-)}(\nu+4\xi){\mathcal S}^{(+)}_\xi(\nu)^{-1}\right]&&\nonumber\\
\times{\cal S}^{(+)}_\xi(\nu){\cal S}^{(+)}_{\xi-1}(\nu)...{\cal S}^{(+)}_{1}(\nu){\bm B}_0(\nu)&=&0;\quad({\rm for}\ \ \xi=1,2,...),\quad\nonumber\\
\label{r1}\\
\left[\Delta_6{\mathcal A}^{(+)}(\nu-4\xi){\mathcal S}^{(-)}_{\xi}(\nu)^{-1}+{\mathcal F}^{-1}(\nu-4\xi)+{\mathcal G}(\nu-4\xi)-{\cal U}+\Delta_6{\mathcal A}^{(-)}(\nu-4\xi){\mathcal S}^{(-)}_{\xi+1}(\nu)\right]&&\nonumber\\
\times{\cal S}^{(-)}_\xi(\nu){\cal S}^{(-)}_{\xi-1}(\nu)...{\cal S}^{(-)}_{1}(\nu){\bm B}_0(\nu)&=&0;\quad({\rm for}\ \ \xi=1,2,...),\quad\nonumber\\
\label{r2}\\
\left[\Delta_6{\mathcal A}^{(+)}(\nu){\mathcal S}^{(+)}_{1}(\nu)+{\mathcal F}^{-1}(\nu)+{\mathcal G}(\nu)-{\cal U}+\Delta_6{\mathcal A}^{(-)}(\nu){\mathcal S}^{(-)}_{1}(\nu)\right]{\bm B}_0(\nu)&=&0.\label{r3}
\end{eqnarray}

The above Eqs. (\ref{r1}, \ref{r2}, \ref{r3}), which are exactly equivalent to the recursion equation (\ref{rbb2}), can be satisfied when the terms in the brackets are zero. By taking the terms in the brackets of  Eqs. (\ref{r1}, \ref{r2}) to be zero, 
we find that ${\mathcal S}^{(\pm)}_\xi(\nu)$ ($\xi=1,2,...$) are given by the continued-fraction-like  recursion equations
 \begin{eqnarray}
{\mathcal S}_\xi^{(\pm)}(\nu)&=&-\Delta_6\frac{1}{\Delta_6{\mathcal A}^{(\pm)}(\nu\pm 4\xi){\mathcal S}^{(\pm)}_{\xi+1}(\nu)+{\mathcal F}^{-1}(\nu\pm 4\xi)+{\mathcal G}(\nu\pm 4\xi)-{\cal U}}{\mathcal A}^{(\mp)}(\nu\pm 4\xi);\quad({\rm for}\ \ \xi=1,2,...),
\label{qpanew}
\end{eqnarray}
with  $\frac{1}{\cal T}$ denoting the inverse of the matrix ${\cal T}$. 

Using the definitions of the matrixes ${\cal F}$, ${\cal G}$ and ${\cal A}^{(\pm)}$, which are shown in our main text, we find that
Eq.~(\ref{qpanew}) implies 
\begin{eqnarray}
\lim_{\xi\rightarrow +\infty}{\mathcal S}_\xi^{(\pm)}(\nu)=-\Delta_6{\mathcal F}(\nu\pm 4\xi){\mathcal A}^{(\mp)}(\nu\pm 4\xi),
\label{qpcnew}
\end{eqnarray}
which is consistent with the requirement of the convergence of the summation in Eq.~(\ref{uex}).
Moreover, Eq.~(\ref{qpanew}) also implies that ${\mathcal S}_\xi^{(+)}(\nu)$ and ${\mathcal S}_\xi^{(-)}(\nu)$ are actually functions of $\nu+4\xi$ and $\nu-4\xi$, respectively. Inspired by these facts, we  formally express ${\mathcal S}_\xi^{(\pm)}(\nu)$ as
\begin{eqnarray}
{\cal S}_\xi^{(\pm)}(\nu)&\equiv&-\Delta_6{\cal Q}^{(\pm)}(\nu\pm4\xi\mp4){\cal F}(\nu\pm 4\xi){\cal A}^{(\mp)}(\nu\pm4\xi);\ \ \ (\xi=1,2,...), \label{bqa2}
\end{eqnarray}
which is just Eq.~(\ref{smm}) of our main text.
Then the continued-fraction-like  recursion equations Eq.~(\ref{qpanew}) and the conditions (\ref{qpcnew}) can be re-expressed as the ones of ${\cal Q}^{(\pm)}(\nu)$, {\it i.e.}, Eqs. (\ref{qp}, \ref{qpcm}) of our main text:
\begin{eqnarray}
{\mathcal Q}^{(\pm)}(\nu)&=&\frac{1}{1+{\cal F}(\nu\pm4)\big[{\cal G}(\nu\pm 4)-{\cal U}\big]-\Delta_6^2 {\cal F}(\nu\pm4) {\cal A}^{(\pm)}(\nu\pm4){\mathcal Q}^{(\pm)}(\nu\pm4){\cal F}(\nu\pm8){\cal A}^{(\mp)}(\nu\pm8)},
\label{qpa}
\end{eqnarray}
and
\begin{eqnarray}
\lim_{z\rightarrow +\infty}{\mathcal Q}^{(+)}(\nu+z)=\lim_{z\rightarrow -\infty}{\mathcal Q}^{(-)}(\nu+z)={\cal I},
\label{qpc}
\end{eqnarray}
with ${\cal I}$ being the $4N\times 4N$ identical matrix. For each given $\nu$, one can  directly calculate the matrix ${\cal Q}^{(\pm)}(\nu)$ via Eqs.~(\ref{qpa}, \ref{qpc}). Some details of this calculation are also given in our main text.

Furthermore, by taking the terms in the brackets of  Eqs.~(\ref{r3}) to be zero and using Eqs.~(\ref{bqa}, \ref{bqa2}), we obtain
\begin{eqnarray}
{\mathcal M}(\nu){\bm B}_0(\nu)&=&0,\label{am}
\end{eqnarray}
where  ${\mathcal M}(\nu)$ is the matrix defined by Eq.~(\ref{mn}) of our main text, {\it i.e.},
\begin{eqnarray}
{\mathcal M}(\nu)
&\equiv&\Delta_6^2 {\cal A}^{(+)}(\nu) {\mathcal Q}^{(+)}(\nu) {\cal F}(\nu+4){\cal A}^{(-)}(\nu+4)-{\cal F}^{-1}(\nu)-{\mathcal G}(\nu)+{\cal U}+\Delta_6^2 {\cal A}^{(-)}(\nu) {\mathcal Q}^{(-)}(\nu)  {\cal F}(\nu-4){\cal A}^{(+)}(\nu-4).\nonumber\\
\label{mnp}
\end{eqnarray}
Eq.~(\ref{am}) yields that $\nu$ is determined by the equation
\begin{eqnarray}
\det[{\mathcal M}(\nu)]=0. \label{nue}
\end{eqnarray}

In summary, with the above discussion we have proved the conclusion that if $\nu$ is a root of the equation (\ref{nue}), then the function ${\bm u}(r)$ given by Eq.~(\ref{uex}), with ${\bm b}_n(\nu)$ being given by Eqs.~(\ref{bb}, \ref{bqa},  \ref{bqa2}, \ref{qpa}, \ref{qpc},   
 \ref{am}), is a solution of Eq.~(\ref{se}). Since Eqs.~(\ref{nue}, \ref{bb}, \ref{bqa},  \ref{bqa2}, \ref{qpa}, \ref{qpc},   
 \ref{am}) are just Eqs.~(\ref{dm}, \ref{bigb}, \ref{bq}, \ref{smm}, \ref{qp}, \ref{qpcm}, \ref{mb}) of our main text, respectively, this conclusion is equivalent to that ${\bm f}_{\epsilon}^{(j)}(r)$ ($j=1,...,N$) of our main text are solutions  of Eq.~(\ref{se}).

In addition, according to Neumann expansion (\ref{uex})  and the recursion equation (\ref{rr}),
if $\{\nu,{\bm b}_n\}$ $(n=0,\pm 1, \pm 2, ...)$ satisfy Eqs.~(\ref{uex}, \ref{rr}), then $\{{\nu}^\prime,{\bm b}_n^\prime\}$ $(n=0,\pm 1, \pm 2, ...)$ also satsify these two equations, if $({\nu}^\prime=-\nu;\ {\bm b}_n^\prime=(-1)^n{\bm b}_{-n})$, or $({\nu}^\prime=\nu+s;\ {\bm b}_n^\prime={\bm b}_{n+s})$ ($s=\pm 1, \pm 2, ...$), or $({\nu}^\prime=\nu^\ast;\ {\bm b}_n^\prime={\rm sign}(\epsilon)^n{\bm b}_{n}^\ast)$, with ${\rm sign}(\epsilon)=1(-1)$ for $\epsilon>0(<0)$. These facts  yield that 
 if $\nu$ is a solution of Eq.~(\ref{nue}), then $-\nu$, $\nu^*$ and $\nu+n$ ($n=0,\pm 1,\pm 2,...$) are also solutions of this equation, as mentioned in the main text.
 
 Moreover, by replacing the  is the first-kind Bessel function $\{J_{\nu+n}(x)|n=0,\pm 1, \pm 2,...\}$  of Eq.~(\ref{uex})
to the  second-kind Bessel function $\{Y_{\nu+n}(x)|n=0,\pm 1, \pm 2,...\}$, and doing the same calculations as above, 
we can prove that ${\bm p}^{(j)}_{\epsilon}(r)\equiv\sqrt{r}\sum_{n=-\infty}^{\infty}{\bm b}_n(\nu_j)Y_{\nu+n}(\sqrt{\bar{\epsilon}}r)$ ($j=1,...,N$), with $\nu_i$ and ${\bm b}_n(\nu_j)$ being given by Eqs.~(\ref{dm}, \ref{bigb}, \ref{bq}, \ref{smm}, \ref{qp}, \ref{qpcm}, \ref{mb}) of our main text,
 are also solutions of Eq.~(\ref{se}), which is linearly independent of ${\bm f}_{\epsilon}^{(j)}(r)$. Due to the relation $Y_{\nu+n}(x)=\left[\cos(\nu\pi)J_{\nu+n}(x)-(-1)^nJ_{-\nu-n}(x)\right]/\sin(\nu\pi)$, the functions ${\bm g}_{\epsilon}^{(j)}(r)$ ($j=1,...,N$) given by Eq.~(\ref{g}) of our main text is a linear combination of ${\bm f}_{\epsilon}^{(j)}(r)$  and
 ${\bm p}^{(j)}_{\epsilon}(r)$, and thus also are solutions  of Eq.~(\ref{se}).

\subsection{B.  Asymptotic Behaviors of ${\bm f}_{\epsilon}^{(j)}(r)$ and ${\bm g}_{\epsilon}^{(j)}(r)$}

\subsubsection{ Asymptotic behaviors for $r\rightarrow 0$.} 

The Bessel function of the first kind can be expressed as
\begin{eqnarray}
J_{\nu}(z)=\sum_{s=0}^{\infty}\frac{(-1)^s}{s!\Gamma(\nu+s+1)}\left(\frac{z}{2}\right)^{2s+\nu}.\label{jv0}
\end{eqnarray}
Substituting this expression into Eq.~(\ref{f}), we can re-express ${\bm f}_{\epsilon}^{(j)}(r)$  as
\begin{eqnarray}
{\bm f}_{\epsilon}^{(j)}(r)&=&\frac{\beta_6^{\frac 12}}{(2y)^{\frac 14}}\sum_{t=0,1,2,3}\sum_{n=-\infty}^{\infty}{\bm b}_{-4n+t}(\nu_j)\sum_{s=0}^{\infty}\frac{(-1)^s\big(\delta_{\epsilon}\Delta_6^{1/4}/2\big)^{2s+\nu_j+t}\left(16/\Delta_6\right)^n}{s!\Gamma(\nu_j-4n+t+s+1)}\left(\frac{y}{2}\right)^{2n-s-\frac{\nu_j+k}{2}},\label{fee}
\end{eqnarray}
with 
\begin{eqnarray}
y=\beta_6^2/(2r^2),
\end{eqnarray}
 and $\delta_\epsilon=1(i)$ for $\epsilon>0$ ($\epsilon<0$). Therefore, the limit $r\rightarrow 0$ is converted into the limit $y\rightarrow +\infty$. As the  in cases with only an isotropic van der Waals potential \cite{Bo1998} or $1/r^3$ potential \cite{Bo1999},
 in the limit $y\rightarrow +\infty$ the right-hand-side of Eq.~(\ref{fee}) is dominated by the terms with 
  with large positive $n$ and small $s$. 
Therefore, 
in this limit we only keep the terms with positive $n$ and $s=0$ \cite{Bo1998, Bo1999}, and thus obtain
\begin{eqnarray}
{\bm f}_{\epsilon}^{(j)}(r\rightarrow 0)&\rightarrow&\frac{\beta_6^{\frac 12}}{(2y)^{\frac 14}}\sum_{t=0,1,2,3}
\sum_{n=0}^{+\infty}{\bm b}_{-4n+t}(\nu_j)\bigg(\frac{\delta_{\epsilon}\Delta_6^{1/4}}{2}\bigg)^{\nu_j+t}\frac{\left(16/\Delta_6\right)^n}{\Gamma(\nu_j-4n+t+1)}\left(\frac{y}{2}\right)^{2n-\frac{\nu_j+t}{2}}\nonumber\\
&\rightarrow&\frac{\beta_6^{\frac 12}}{(2y)^{\frac 14}}\sum_{t=0,1,2,3}{\bm q}^{(-)}_t(\nu_j)\sum_{n=0}^{+\infty}\frac{(-1)^n}{\Gamma^2[n+1-(\nu_j+t)/4]}\left(\frac{y}{2}\right)^{2n-\frac{\nu_j+t}{2}},
\label{f02}
\end{eqnarray}
with
\begin{eqnarray}
{\bm q}^{(-)}_t(\nu_j)\equiv\lim_{n\rightarrow\infty}\Bigg\{{\bm b}_{-4n+t}(\nu_j)\frac{\big({\rm-}16/\Delta_6\big)^n\Gamma^2\big[n+1-(\nu_j+t)/4\big]}{\big(\delta_\epsilon \Delta_6^{1/4}/2\big)^{-(\nu_j+t)}\Gamma(\nu_j-4n+t+1)}\Bigg\};\ \ (t=0,1,2,3).
\end{eqnarray}
Using the property of Gamma function 
\begin{eqnarray}
\lim_{z\rightarrow+\infty}\frac{\Gamma(z+a)\Gamma(z-a)}{\Gamma(z)\Gamma(z)}=1,
\end{eqnarray}
and comparing the summation on the second line in Eq.~(\ref{f02}) with series of Bessel function in Eq.~(\ref{jv0}), we finally get
\begin{eqnarray}
{\bm f}_{\epsilon}^{(j)}(r\rightarrow 0)\rightarrow\sqrt{r}\sum_{t=0,1,2,3}{\bm q}^{(-)}_k(\nu_j)J_{-\frac{\nu_j+t}{2}}\bigg(\frac{\beta_6^2}{2r^2}\bigg).\label{fap}
\end{eqnarray}
Substituting the fact
\begin{eqnarray}
J_{\nu}\big(|z|\rightarrow +\infty\big)\rightarrow\sqrt{\frac{2}{\pi z}}\cos\left(z-\frac{\pi\nu}{2}-\frac{\pi}{4}\right);\quad(-\pi<\arg z<\pi).\label{jvi}
\end{eqnarray}
into Eq.~(\ref{fap}), we finally obtain Eqs.~(\ref{f0}, \ref{pcs}, \ref{qk}) of our main text. 

The asymptotic behaviors of ${\bm g}_{\epsilon}^{(j)}(r)$ for $r\rightarrow 0$, {\it i.e.}, Eqs.~(\ref{g0}, \ref{pcs}, \ref{qk}) of our main text, can be obtained with the above approach.

\subsubsection{Asymptotic behaviors for $r\rightarrow\infty$.} 

For $\epsilon>0$, the asymptotic behavior of ${\bm f}_{\epsilon}^{(j)}(r)$ and  ${\bm g}_{\epsilon}^{(j)}(r)$ in the limit $r\rightarrow\infty$, {\it i.e.}, Eqs.~(\ref{fi1}-\ref{dcs}) of our main text, can be directly obtained by substituting Eq.~(\ref{jvi}) into Eqs.~(\ref{f}, \ref{g}) of our main text.

For $\epsilon<0$, we prove the result (\ref{hrl}) of our main text can  by substituting the relations $H_\eta^{(1)}(z)=\left[J_{-\eta}(z)-e^{-i\eta\pi}J_{\eta}(z)\right]/[i\sin(\eta\pi)]$ and $H_\eta^{(2)}(z)=\left[J_{-\eta}(z)-e^{i\eta\pi}J_{\eta}(z)\right]/[-i\sin(\eta\pi)]$, with $H_\eta^{(1,2)}$ being the first and second kind of Hankel functions, into Eqs.~(\ref{f}, \ref{g}) and then using the definition of ${\bm h}_{\epsilon}^{(j)}(r)$ in our main text, and using the fact that $\lim_{z\rightarrow+\infty}H_\eta^{(1)}(iz)\rightarrow(-i)e^{-i\eta\pi/2}\sqrt{2/(\pi z)}e^{-z}$.

\subsection{C. Solutions with Energy-Independent Boundary Conditions for $r\rightarrow 0$}

We define the solutions ${\bm v}^{(j)}_{\epsilon}(r)$ and ${\bm w}^{(j)}_{\epsilon}(r)$ ($j=1,...,N$) of Eq.~(\ref{se}) as
\begin{eqnarray}
{\bm v}^{(j)}_{\epsilon}(r)&=&\sum_{t=1}^{N}{\bm f}_{\epsilon}^{(t)}(r)V_{tj}+\sum_{t=1}^{N}{\bm g}_{\epsilon}^{(t)}(r)V^\prime_{tj};\label{vd}\\
{\bm w}^{(j)}_{\epsilon}(r)&=&\sum_{t=1}^{N}{\bm f}_{\epsilon}^{(t)}(r)W_{tj}+\sum_{t=1}^{N}{\bm g}_{\epsilon}^{(t)}(r)W^\prime_{tj},\ \ \  (j=1,...,N),\label{wd}
\end{eqnarray}
where the coefficients  $V_{tj}$, $V^\prime_{tj}$, $W_{tj}$ and $W^\prime_{tj}$ ($t,j=1,...,N$) are given by
\begin{eqnarray}
{\mathbb V}&=&\frac{1}{{\mathbb P}_{{\rm c}}^{(-)}-{\mathbb P}_{{\rm c}}^{(+)}\frac{ \displaystyle 1}{
\displaystyle
 {\mathbb P}_{{\rm s}}^{(+)}
 }{\mathbb P}_{{\rm s}}^{(-)}};
 \hspace{0.6cm}
 {\mathbb V}^{\prime}=-
 \frac{1}{\displaystyle
 {\mathbb P}_{{\rm s}}^{(+)}}
 {\mathbb P}_{{\rm s}}^{(-)}
 \frac{ \displaystyle 1}{{\mathbb P}_{{\rm c}}^{(-)}-{\mathbb P}_{{\rm c}}^{(+)}\frac{\displaystyle 1}{
\displaystyle
 {\mathbb P}_{{\rm s}}^{(+)}
 }{\mathbb P}_{{\rm s}}^{(-)}};\label{vv2}\\
 {\mathbb W}&=&\frac{1}{{\mathbb P}_{{\rm s}}^{(-)}-{\mathbb P}_{{\rm s}}^{(+)}\frac{ \displaystyle 1}{
\displaystyle
 {\mathbb P}_{{\rm c}}^{(+)}
 }{\mathbb P}_{{\rm c}}^{(-)}};
 \hspace{0.6cm}
 {\mathbb W}^{\prime}=-
 \frac{1}{\displaystyle
 {\mathbb P}_{{\rm c}}^{(+)}}
 {\mathbb P}_{{\rm c}}^{(-)}
 \frac{ \displaystyle 1}{{\mathbb P}_{{\rm s}}^{(-)}-{\mathbb P}_{{\rm s}}^{(+)}\frac{\displaystyle 1}{
\displaystyle
 {\mathbb P}_{{\rm c}}^{(+)}
 }{\mathbb P}_{{\rm c}}^{(-)}}.\label{ww2}
\end{eqnarray}
Here ${\mathbb V}$, ${\mathbb V}^{\prime}$, ${\mathbb W}$ and ${\mathbb W}^{\prime}$ are the $N\times N$ matrixes with elements $V_{tj}$, $V^\prime_{tj}$, $W_{tj}$ and $W^\prime_{tj}$ ($t,j=1,...,N$), respectively, and the $N\times N$ matrixes ${\mathbb P}_{{\rm c}}^{(\pm)}$ and ${\mathbb P}_{{\rm s}}^{(\pm)}$  are defined as
\begin{eqnarray}
{\mathbb P}_{{\rm c}}^{(\pm)}&=&\big[{\bm p}_{{\rm c}\pm}(\nu_1),\  {\bm p}_{{\rm c}\pm}(\nu_2), \ ...,\  {\bm p}_{{\rm c}\pm}(\nu_N)\big];\\
{\mathbb P}_{{\rm s}}^{(\pm)}&=&\big[{\bm p}_{{\rm s}\pm}(\nu_1),\  {\bm p}_{{\rm s}\pm}(\nu_2), \ ...,\  {\bm p}_{{\rm s}\pm}(\nu_N)\big],
\end{eqnarray}
where ${\bm p}_{{\rm c}\pm}(\nu_j)$ and ${\bm p}_{{\rm s}\pm}(\nu_j)$ ($j=1,...,N$) are defined in Eq.~(\ref{pcs}) of our main text.

In the limit $r\rightarrow 0$, ${\bm v}^{(j)}_{\epsilon}(r)$ and ${\bm w}^{(j)}_{\epsilon}(r)$ ($j=1,...,N$) have energy-independent asymptotic behaviors:
\begin{eqnarray}
{\bm v}^{(j)}_{\epsilon}(r\rightarrow 0)&=&\frac{2r^{3/2}}{\sqrt{\pi}\beta_6}{\bm \delta}_j
\cos\left(\frac{\beta_6^2}{2r^2}-\frac{\pi}{4}\right)+{\cal O}(r^{5/2});\\
{\bm w}^{(j)}_{\epsilon}(r\rightarrow 0)&=&\frac{2r^{3/2}}{\sqrt{\pi}\beta_6}{\bm \delta}_j
\sin\left(\frac{\beta_6^2}{2r^2}-\frac{\pi}{4}\right)+{\cal O}(r^{5/2}),
\end{eqnarray}
where ${\bm \delta}_j$ ($j=1,...,N$) is the $N$-component vector with the $j$-th component being 1 and other components being zero. Namely, in the limit $r\rightarrow 0$, the $j$-th component of ${\bm v}^{(j)}_{\epsilon}$ and ${\bm w}^{(j)}_{\epsilon}$ tend to $\frac{2r^{3/2}}{\sqrt{\pi}\beta_6}
\cos\left(\frac{\beta_6^2}{2r^2}-\frac{\pi}{4}\right)+{\cal O}(r^{5/2})$ and 
$\frac{2r^{3/2}}{\sqrt{\pi}\beta_6}
\sin\left(\frac{\beta_6^2}{2r^2}-\frac{\pi}{4}\right)+{\cal O}(r^{5/2})$, respectively, while all other components of ${\bm v}^{(j)}_{\epsilon}$ and ${\bm w}^{(j)}_{\epsilon}$  tend to ${\cal O}(r^{5/2})$ ($j=1,...,N$). 

Furthermore, the behaviors of  ${\bm v}^{(j)}_{\epsilon}(r)$ and ${\bm w}^{(j)}_{\epsilon}(r)$ ($j=1,...,N$) in the long-range limit $r\rightarrow \infty$  can be directly derived via Eqs.~(\ref{vd}, \ref{wd}) and Eqs.~(\ref{fi1}, \ref{gi1}) of our main text as
\begin{eqnarray}
{\bm v}^{(j)}_{\epsilon}(r\rightarrow\infty)&\rightarrow&\sqrt{\frac{2}{\pi\sqrt{\bar \epsilon}}}\bigg[\widetilde{\bm v}^{(j)}\cos\left(\sqrt{\bar \epsilon}r\right)-\widetilde{\bm v}'^{(j)}\sin\left(\sqrt{\bar \epsilon}r\right)\bigg];\\
{\bm w}^{(j)}_{\epsilon}(r\rightarrow\infty)&\rightarrow&\sqrt{\frac{2}{\pi\sqrt{\bar \epsilon}}}\bigg[\widetilde{\bm w}^{(j)}\cos\left(\sqrt{\bar \epsilon}r\right)-\widetilde{\bm w}'^{(j)}\sin\left(\sqrt{\bar \epsilon}r\right)\bigg].
\end{eqnarray}
Here the $N$-component vectors $\widetilde{\bm v}^{(j)}$, $\widetilde{\bm v}'^{(j)}$, $\widetilde{\bm w}^{(j)}$ and $\widetilde{\bm w}'^{(j)}$ ($j=1,...,N$) are given by the relations of the $N\times N$ matrixes
\begin{eqnarray}
\left[\widetilde{\bm v}^{(1)}, \widetilde{\bm v}^{(2)},...,\widetilde{\bm v}^{(N)}\right]&=&{\mathbb D}_{{\rm c}}^{(-)}{\mathbb V}+{\mathbb D}_{{\rm c}}^{(+)}{\mathbb V}';\\
\left[\widetilde{\bm v}'^{(1)}, \widetilde{\bm v}'^{(2)},...,\widetilde{\bm v}'^{(N)}\right]
&=&{\mathbb D}_{{\rm s}}^{(-)}{\mathbb V}+{\mathbb D}_{{\rm s}}^{(+)}{\mathbb V}';\\
\left[\widetilde{\bm w}^{(1)}, \widetilde{\bm w}^{(2)},...,\widetilde{\bm w}^{(N)}\right]
&=&{\mathbb D}_{{\rm c}}^{(-)}{\mathbb W}+{\mathbb D}_{{\rm c}}^{(+)}{\mathbb W}';\\
\left[\widetilde{\bm w}'^{(1)}, \widetilde{\bm w}'^{(2)},...,\widetilde{\bm w}'^{(N)}\right]
&=&{\mathbb D}_{{\rm s}}^{(-)}{\mathbb W}+{\mathbb D}_{{\rm s}}^{(+)}{\mathbb W}',
\end{eqnarray}
with the $N\times N$ matrixes ${\mathbb V}$, ${\mathbb V}^\prime$, ${\mathbb W}$ and ${\mathbb W}^\prime$ being defined in Eqs.~(\ref{vv2}, \ref{ww2}), and ${\mathbb D}_{{\rm c}}^{(\pm)}$ and ${\mathbb D}_{{\rm s}}^{(\pm)}$  being defined as
\begin{eqnarray}
{\mathbb D}_{{\rm c}}^{(\pm)}&=&\big[{\bm d}_{{\rm c}\pm}(\nu_1),\  {\bm d}_{{\rm c}\pm}(\nu_2), \ ...,\  {\bm d}_{{\rm c}\pm}(\nu_N)\big];\\
{\mathbb D}_{{\rm s}}^{(\pm)}&=&\big[{\bm d}_{{\rm s}\pm}(\nu_1),\  {\bm d}_{{\rm s}\pm}(\nu_2), \ ...,\  {\bm d}_{{\rm s}\pm}(\nu_N)\big],
\end{eqnarray}
where ${\bm d}_{{\rm c}\pm}(\nu_j)$ and ${\bm d}_{{\rm s}\pm}(\nu_j)$ ($j=1,...,N$) are defined in Eq.~(\ref{dcs}) of our main text.

\subsection{D. Application of Our Method to the General Cases}

In this section we show how to apply our method shown in our main text and Sec.~A, B of this supplementary material to derive the solutions of the Schr\"odinger equation 
\begin{eqnarray}
\left[-\frac{\hbar^2\nabla_{\bm r}^2}{2\mu}+\sum_{\lambda= 2}^{\lambda_{\rm max}} \frac{V_\lambda(\theta,\varphi)}{r^\lambda}\right]\Psi({\bm r})=\epsilon\Psi({\bm r}),\label{se2}
\end{eqnarray}
which is projected to the subspace with $l=0,1,2,...,l_{\rm cut}$, with $\epsilon\neq 0$. 
We assume that the interaction term of Eq.~(\ref{se2}) can couple the eigen-sates of angular-momentum $L^2$ and $L_z$ with quantum numbers $(l_1,m_1)$, $(l_2,m_2)$,..., $(l_N,m_N)$. Therefore, as in our main text, we expand the solution of Eq.~(\ref{se2}) as
\begin{eqnarray}
\Psi({\bm r})=\sum_{\alpha=1}^{N}\frac{u_{l_\alpha,m_\alpha}(r)}{r}{Y}_{l_\alpha}^{m_\alpha}(\theta,\varphi).\label{psi2}
\end{eqnarray}
We further
define the the $N$-component radial wave function ${\bm u}(r)$ as
\begin{eqnarray}
{\bm u}(r)
\equiv
\begin{pmatrix}
u_{l_1,m_1}(r)\\
u_{l_2,m_2}(r)\\
...\\
u_{l_N,m_N}(r)
\end{pmatrix}.\label{ucc2}
\end{eqnarray}
Accordingly,  Eq.~(\ref{se2})
 can be re-written as the equation for ${\bm u}(r)$:
\begin{eqnarray}
\left[\frac{\textrm{d}^2}{\textrm{d}r^2}-\frac{{\mathbb C}}{r^2}-\sum_{\lambda=2}^{\lambda_{\rm  max}}\frac{{\mathbb V}^{(\lambda)}}{r^\lambda}+\bar{\epsilon}\hspace{0.05cm} {\mathbb I}\right]{\bm u}(r)=0,\label{se2}
\end{eqnarray}
where ${\mathbb C}$ and ${\mathbb V}^{(\lambda)}$ ($\lambda=3,...,\lambda_{\rm max}$) are $N\times N$ matrixes with elements
\begin{eqnarray}
&&{\mathbb C}_{ij}=l_i(l_i+1)\delta_{ij};
\hspace{0.6cm}
{\mathbb V}^{(\lambda)}_{ij}=\frac{2\mu}{\hbar^2}\int_0^\pi d\theta \int_0^{2\pi} d\phi \sin\theta Y_{l_i}^{m_i\ast}(\theta,\varphi)V_\lambda(\theta,\varphi) Y_{l_j}^{m_j}(\theta,\varphi),\ \ \ (i,j=1,2,...,N). 
\end{eqnarray}

As in Sec.~A of this supplementary material, we expand the solution ${\bm u}(r)$ of Eq.~(\ref{se2})  as a
 Neumann series:
\begin{eqnarray}
{\bm u}(r)=\sqrt{r}\sum_{n=-\infty}^{\infty}{\bm b}_n(\nu)J_{\nu+n}(x),\label{uex2}
\end{eqnarray}
where $x=\sqrt{\bar{\epsilon}}r$.
We substitute
 the form (\ref{uex2}) into Eq.~(\ref{se2}) and then obtain 
\begin{eqnarray}
\left[x^2\frac{\textrm{d}^2}{\textrm{d}x^2}{\mathbb I}+x\frac{\textrm{d}}{\textrm{d}x}{\mathbb I}+x^2{\mathbb I}-{\mathbb C}-\frac{1}{4}{\mathbb I}-\sum_{\lambda=2}^{\lambda_{\rm max}}\left(\frac{\sqrt{\bar{\epsilon}}}{x}\right)^{\lambda-2}{\mathbb V}^{(\lambda)}\right]\Bigg\{\sum_{n=-\infty}^{\infty}{\bm b}_n(\nu)J_{\nu+n}(x)\Bigg\}=0,\label{5e}
\end{eqnarray}

Substituting the property (\ref{jp}) of Bessel function
into Eq.~(\ref{5e}), we further derive the 
the recursion equation of ${\bm b}_n(\nu)$, which has the form of Eq.~(\ref{rr}). In this  recursion equation,  ${\bm b}_n(\nu)$ is coupled to ${\bm b}_{n\pm 1}(\nu)$, ${\bm b}_{n\pm 2}(\nu)$, ..., ${\bm b}_{n\pm n_\ast}(\nu)$,
with
\begin{eqnarray}
n_\ast=\lambda_{\rm max}-2. 
\end{eqnarray}
Therefore, as in Sec.~A of this supplementary material, we can combine ${\bm b}_{n}$ and  ${\bm b}_{n+1}$, ${\bm b}_{n+2}$,..., ${\bm b}_{n+n_\ast-1}$ together,
and introduce the $(N n_\ast)$-component vector  ${\bm B}_\alpha(\nu)$ which is defined as
\begin{eqnarray}
{\bm B}_\alpha(\nu)\equiv
\begin{bmatrix}
{\bm b}_{n_\ast\alpha+n_\ast-1}(\nu)\\
...\\
 {\bm b}_{n_\ast\alpha+2}(\nu) \\
{\bm b}_{n_\ast\alpha+1}(\nu)\\	
{\bm b}_{n_\ast\alpha}(\nu)
\end{bmatrix},
\ \ (\alpha=0,\pm 1,\pm 2,...).\label{bb2}
\end{eqnarray}
Then the 
the recursion equation of ${\bm b}_n(\nu)$ can be re-written as the recursion equation of ${\bm B}_\alpha(\nu)$, where ${\bm B}_\alpha(\nu)$ is only coupled to ${\bm B}_{\alpha\pm 1}(\nu)$, like Eq.~(\ref{rbb2}). Then we can directly use the approaches shown in Sec.~A, Sec.~B and our main text to derive the solutions of Eq.~(\ref{se2}).

\end{document}